\documentclass[journal,onecolumn, draftcls, 12pt]{IEEEtran} % for arxiv

\usepackage[utf8]{inputenc}
\usepackage[T1]{fontenc}
\usepackage[hidelinks]{hyperref}
\usepackage{url}
\usepackage{ifthen}
\usepackage{cite}
\usepackage[cmex10]{amsmath} % Use the [cmex10] option to ensure complicance
\usepackage{amsfonts,amssymb}
\usepackage{color}
\usepackage{ifthen}
\usepackage[arxiv]{optional} % 2col/arxiv

\newcommand{\eq}[1]{\begin{align}#1\end{align}}

\newcommand{\lb}[1]{\left\{ \begin{array}{ll} #1 \end{array} \right.}

\newcommand{\pushright}[1]{\ifmeasuring@ #1 \else\omit\hfill$\displaystyle#1$\fi\ignorespaces}
\newcommand{\pushleft}[1]{\ifmeasuring@ #1 \else\omit$\displaystyle#1$\hfill\fi\ignorespaces}

\newtheorem{lemma}{Lemma}
\newtheorem{corollary}{Corollary}

\newtheorem{theorem}{Theorem}

\newtheorem{example}{Example}

\newcommand{\E}{\mathbb{E}}

\begin{document}
\title{Fault Tolerant Equilibria in Anonymous Games: best response correspondences and fixed points}
\author{
 \IEEEauthorblockN{Deepanshu Vasal and Randall Berry}\\
  \IEEEauthorblockA{ \small{{dvasal@umich.edu, rberry@northwestern.edu}}}
 }

\maketitle

%%%%%%
%% Abstract:
%% If your paper is eligible for the student paper award, please add
%% the comment "THIS PAPER IS ELIGIBLE FOR THE STUDENT PAPER
%% AWARD." as a first line in the abstract.
%% For the final version of the accepted paper, please do not forget
%% to remove this comment!
%%
\begin{abstract}
The notion of {\it fault tolerant Nash equilibria} has been introduced as a way of studying the robustness of Nash equilibria. Under this notion, a fixed number of players are allowed to exhibit faulty behavior in which they may deviate arbitrarily from an equilibrium strategy. A Nash equilibrium in a game with $N$ players is said to be $\alpha$-tolerant if no non-faulty user wants to deviate from an equilibrium strategy as long as $N-\alpha-1$ other players are playing the equilibrium strategies, i.e., it is robust to deviations from rationality by $\alpha$ faulty players. In prior work, $\alpha$-tolerance has been largely viewed as a property of a given Nash equilibria. Here, instead we consider following Nash's approach for showing the existence of equilibria, namely, through the use of best response correspondences and fixed-point arguments. In this manner, we provide sufficient conditions for the existence an $\alpha$-tolerant equilibrium. This involves first defining an {\it $\alpha$-tolerant best response correspondence}. Given a strategy profile of non-faulty agents, this  correspondence contains strategies for a non-faulty player that are a best response given {\it any} strategy profile of the faulty players. We prove that if this correspondence is non-empty, then it is upper-hemi-continuous. This enables us to apply Kakutani's fixed-point theorem and  argue that if this correspondence is non-empty for every strategy profile of the non-faulty players then there exists an $\alpha$-tolerant equilibrium. However, we also illustrate by examples, that in many games this best response correspondence will be empty for some strategy profiles even though $\alpha$-tolerant equilibira still exist.
\end{abstract}

%\vspace*{-0.15cm}
\section{Introduction}

Nash equilibrium has become one of the most common solution concepts used to study (normal form) games.  However, this concept it is not without its critics. Indeed, many researchers have pointed out short-comings of Nash equilibria and sought to address these, e.g.,  through various equilibrium refinements. In this paper, we focus on one such short-coming, namely, the robustness of Nash equilibrium to deviations of some players from the prescribed equilibrium behavior, and one approach to address this, namely, the use of {\it fault tolerant Nash equilibrium} (e.g.~\cite {Ha11,Ka18,GK21}).  A Nash equilibrium is said to be {\it $\alpha$-tolerant}, if that 
equilibrium is robust to deviations by $\alpha$ faulty players, where a faulty player is one that may chose any arbitrary action.  In other words, in a game with $N$ players, no (non-faulty) player in a $\alpha$-tolerant equilibrium would want to deviate from the equilibrium as long as $N-\alpha -1$ players continued to play the equilibrium strategies.  This models situations in which some subset of players may behave in unexpected ways, which may be motivated by modeling errors, imperfect rationality of other players, or effects not included within the model. 
  This notion of fault tolerance has a long history of use in distributed computing, where the focus is on designing algorithms that are implemented by a set of distributed processors. The goal is to ensure that the given algorithm is robust to faults by some number of these processors (e.g.~\cite{FLP85}).  Here, instead the faults are incurred by players in a game taking actions that are not expected at a Nash equilibrium. 
  
One reason for the wide-spread use of Nash equilibria rests on Nash's celebrated result showing the existence of Nash equilibria in finite games \cite{Na50}.  This can be shown by applying Kakutani's fixed point theorem \cite{Ka41} to the player's best-response correspondences in the game.  For a given $\alpha$, it is easy to construct games for which no $\alpha$-tolerant equilibrium exists, which shows that Nash's existence result can not be directly generalized for $\alpha$-tolerant Nash equilibria.  In this paper, we follow the approach of studying the existence of $\alpha$-tolerant Nash equilibria via studying best response correspondences.  We define a natural {\it $\alpha$-tolerant best response correspondence} and show that if this correspondence is non-empty for every strategy profile of the non-faulty players, then one can again appeal to Kakutani's fixed point theorem to show that a $\alpha$-tolerant Nash equilibrium exists.  However, we also show via a sequence of examples that the requirement that the best response correspondence is non-empty for every strategy profile is often not satisfied even in games that have equilibria which are tolerant to a large number of faults.  The main issue here is that away from the equilibrium, the faulty agents can have a large impact on the agents' best responses.  This suggest that the existence of fault-tolerant equilibria is often determined by the ``local" behavior of the best response correspondence near the equilibria and not global properties of this map. 

Our analysis in this paper is focused on {\it anonymous games} \cite{Blo99} with finite action spaces and symmetric pay-offs. These are a class of games in which each player's pay-off does not depend on the identity of the other players, but only on the number of each player choosing a particular action. As pointed out in \cite{Blo99} such games arise in many economic models. Further, this restriction is analytically convenient for us as  as one can simply look at a fixed set of $\alpha$ players as deviating from the equilibrium, while in a more general game the impact of $\alpha$ players deviating may vary depending on which set of players deviates.  

In terms of related work, an early paper to use idea of fault tolerance in games is the work in \cite{El02}, which considered fault tolerance in 
an implementation setting. A case for combining fault tolerance with Nash equilibrium was also discussed in \cite{Ha11, Ha03}.
In \cite{GR14}, existence results where given for $(\epsilon,\alpha)$-tolerant Nash equilibria for certain classes of 
``large" games, including anonymous games. In an $(\epsilon,\alpha)$-tolerant equilibria, players will only deviate if they can improve 
their pay-off by at least $\epsilon$. Here, we focus on the case where such an $\epsilon$ margin does not exist (i.e.~$\epsilon =0$),
which is not covered by the results in \cite{GR14}.

In \cite{Ka18}, a defection-deterrence index of an equilibrium is studied, which characterizes the smallest number of faulty agents that can 
result in other agents seeking to deviate from the equilibrium.  In other words, if an equilibrium has a defection index of $d$, then 
it is $d-1$-tolerant, but not $d$-tolerant.  \cite{Ka18} also introduced a formation index of an equilibria, which given the smallest number of players $f$ such that if all $f$ players commit to a given equilibrium, then playing that equilibrium is a best response for the remaining players.  
Further, it is shown that these two indices are dual in that their sum is always equal to the number of players in the game.  

This work also has ties to the vast literature on robustness of Nash equilibria and various refinements. We 
highlight only a few of these in the following. First, this has connections to the work on ``robust Nash equilibrium" (e.g.~\cite{AB04,HYF05}) that uses ideas from robust optimization \cite{BN02}.
In this work, players can be viewed as taking a worst-case view over any uncertainty.  For a model with faulty agents this would correspond to assuming that that faulty agents always choose the action profile that minimizes a players pay-off, similar to the classic  max-min solutions used for zero-sum games \cite{NeMo44}. It can be seen that a $\alpha$-tolerant Nash equilibrium is also a robust Nash equilibrium in this sense. However, the converse does not hold. A game may have robust Nash equilibrium that are not $\alpha$-tolerant, i.e., $\alpha$-tolerance is a stronger equilibrium notion.  This also has ties to work on coalition proof Nash equilibrium \cite{BPW87} or strong Nash equilibrium \cite{Au59}, which allow for coalitions of agents to deviate from a proposed equilibrium if this benefits the agents in the coalition.  These notions allow for arbitrary sets of agents to form coalitions to deviate, while here we fix the number of faulty agents at $\alpha$.\footnote{In \cite{GR14}, $t$-coalitional Nash equilibria are studied, which restricts attentions to deviations by coalitions of size $t$.  In \cite{GR14} it is shown that this can be related $\alpha$-tolerance and a related notion of $\alpha$-immunity.} Additionally, the deviations in these notions require that it benefit the agents in the deviating coalition, while here we allow for arbitrary deviations. Another line of work is that using the notion of ``robustness to incomplete information" as given in \cite{KM97} (see also \cite{Ui01}). This line work captures faulty behavior by considering  games of incomplete information that are "near" an underlying game of complete information.  In these nearby games, some players have a dominant strategy to choose one action instead of their best response in the complete information game and thus can be viewed as faulty. This notion assumes that  these deviations from the complete information game occur with low probability, while in a $\alpha$-tolerant equilibrium, no probabilistic assumptions about the deviations are made.

%We note that for $\alpha  = 0$, this boils down to the standard Nash equilibrium which exists for any finite game and for $\alpha = N$, it becomes dominant strategy equilibrium.
%We refer the reader to~\cite{Ka18} for numerous examples of games for which $\alpha$-robust equilibria exist when 
%$\alpha >0$. It is also easy to construct examples of finite games for which $\alpha$-robust equilibria do not exist for any $\alpha>0$. For a given $\alpha >0$, a fundamental question is then to provide conditions for when a $\alpha$-robust equilibrium exists. This paper addresses this question. Namely, we provide provide sufficient conditions for the existence of a $\alpha$-robust equilibrium in  {\it anonymous games} \cite{Blo99} with finite action spaces and symmetric pay-offs. These are a class of games in which each player's pay-off does not depend on the identity of other players, but only on the number of each player chosing a particular action. As pointed out in \cite{Blo99} such games arise in many economic models and are a natural setting to study this equilibrium concept as one can simply look at a fixed set of $\alpha$ players as deviating from the equilibrium, while in a more general game the impact of $\alpha$ players deviating may vary depending on which set of players deviates.

The remainder of the paper is structured is as follows. In Section~\ref{sec:Model}, we define our model and give our definition of  
$\alpha$-tolerant Nash equilibrium. In Section~\ref{sec:Exist}, we introduce the notion of a $\alpha$-tolerant best response and 
use this to provide sufficient conditions for the existence of the equilibrium. We also present example showing when this condition may not be satisfied.
%In Section~\ref{sec:Example}, we provide an example. 
We conclude in Section~\ref{sec:Conc}.

\section{Model and Equilibrium Definitions}
\label{sec:Model}
%In this paper, we consider both finite and infinite player anonymous games where..... 

%\subsection{Finite number of players}

\subsection{Anonymous Game Model}
We first formally define the class of symmetric anonymous games that we will focus on in this paper. 
Consider a game $G(\mathcal{N},\mathcal{A},u)$, where $\mathcal{N}$ is a set of $N$ players, each with the same finite action space $\mathcal{A}$ and utility function $u$.  Each player $i$'s utility is given by
\eq{
u(a^i, f(a^{-i}))
}
where for all $i$, $a^i\in\mathcal{A}$ is the action of player $i$, $a^{-i}$ is the set of actions of all players other than $i$, and $f(a^{-i})$ is the frequency distribution of $a^{-i}$, i.e., this specifies the number of times each action in $\mathcal{A}$ appears in $a^{-i}$. 
This form of utility captures the anonymous nature of these games in that a player's utility only depends on the distribution of actions of the other players and not on the specific players. Further, these games are symmetric in that all players have the same utility function and action space.
 %i.e. $f(a^{-i})= f(\pi(a^{-i}))$, where $\pi:\mathcal{A}^{N-1}\to \mathcal{A}^{N-1}$ is the random permutation function. Some examples of such an $f$ include $f(a^{-i}) = \max^{(n)}(a^{-i})$, where $\max^{(n)}$ is the $n^{th}$ order statistic, or $f(a^{-i})=\sum_{j\neq i}a^j$.

Let $\Delta(\mathcal{A})$ denote the set of probability distributions over $\mathcal{A}$.
A (mixed) strategy profile of the game is given by  $\underline{\sigma} = (\sigma^i)_{i \in \mathcal{N}}$, where for each $i$, $\sigma^i \in\Delta(\mathcal{A})$ is a mixed strategy for player $i$.  

Given a mixed strategy profile, $\underline{\sigma}$, the utility obtained by player $i$ is denoted by 
\eq{
U_i(\underline{\sigma}) =  \E^{\underline{\sigma}} u(A^i, f(A^{-i}))
}
where $A^i$ denotes a random variable distributed according to $\sigma^i$, $A^{-i}$ denotes a set of $N-1$ random variables distributed according to the product measure $\prod_{j\neq i} \sigma^j$, and the expectation is taken with respect to these random variables.

A mixed strategy profile $\underline{\sigma}^*$ is a {\it Nash equilibrium} of $G$ if 
\eq{\label{eq:NE}
U_i(\underline{\sigma}^*) \geq U_i(\sigma^i,\underline{\sigma}^{-i,*})
}
for all $i$ and all $\sigma^i \in \Delta(\mathcal{A}),$ where $\underline{\sigma}^{-i,*}$ denotes 
the set of mixed strategies in $\underline{\sigma}^*$ for all players except $i$. 

Note that in (\ref{eq:NE}), it is assumed that all other players except $i$ are playing the given equilibrium strategy.
The notion of fault-tolerance, introduced next, relaxes this assumption.

\subsection{Fault-tolerant Nash Equilibrium}
\label{sec:EqConcept}
%\subsection{Finite Users}

For a given $\alpha \in \{0,1,\ldots, N\}$, divide the set of $N$ players into a set of $\alpha$ {\it faulty} players 
and a set of $N-\alpha$ {\it normal} or {\it non-faulty} players. Without loss of generality, due to the symmetric and anonymous nature of the game, we assume that players 
$1,\ldots, N-\alpha$ are normal and the remaining players are faulty. We allow the faulty players to choose any arbitrary mixed strategy in $\Delta(\mathcal{A})$. Let $\underline{\tau}^\alpha = (\tau_i)_{i=N-\alpha +1}^{N}$ denote such a mixed strategy profile for the $\alpha$ faulty players, where
$\tau_i \in \Delta({\mathcal A})$ for all $i = N-\alpha+1,\ldots,N$ and let $T^\alpha$ denote the set of all such profiles.
Likewise, let $\underline{\sigma}^{(\alpha)}=\{\sigma^{k,(\alpha)}\}_{k= 1}^{N-\alpha}$ denote a mixed strategy profile for the $N-\alpha$ non-faulty players.

The mixed strategy profile  $\underline{\sigma}^{*(\alpha)}$ is a {\it $\alpha$-tolerant Nash equilibrium} 
if 
\eq{\label{eq:tol-NE}
U_i(\underline{\sigma}^{*(\alpha)},\underline{\tau}^{\alpha}) \geq U_i(\sigma^i,\underline{\sigma}^{-i,*(\alpha)},\underline{\tau}^{\alpha})
}
for all $i = 1,\ldots,N-\alpha$, all $\sigma^i \in \Delta(\mathcal A)$, and every $\underline{\tau}^{\alpha} \in T^{\alpha}$.  

In other words, every non-faulty player can not unilaterally improve their pay-off regardless of the strategy profile of the faulty agents.

Note that a $0$-tolerant Nash equilibrium corresponds to a normal Nash equilibrium of a game as in this case there are no faulty agents.
The following gives an example of a game that has a $\alpha$-tolerant Nash equilibrium for $\alpha >0$.

\begin{example}[A popularity game]\label{ex:1}
{\it 
Consider an $N$ player game of matching actions, where each player chooses an action from the set $\{1,2,3\}$, such that a player's utility is 1 if it plays an action that is played by maximum number of people. Thus 
\eq{
u(a^i,f(a^{-i})) &= \lb{1 \text{  if  } a^i \in  \arg\max f(a^{-i})\\
0 \text{ otherwise }. 
}
}
Here, $a \in \arg\max f(a^{-i})$ denotes that no other action appears more frequently than $a$ in $a^{-i}$.
Clearly all agents playing the same action $a\in\{1,2,3\}$ is a pure strategy Nash equilibria. Suppose now that there are $\alpha$-faulty players.  
All of the non-faulty agents playing the same action $a\in\{1,2,3\}$ is an $\alpha$-tolerant Nash 
equilibrium, if the number of other non-faulty agents, $N-1-\alpha$, is larger than the number of faulty agents $\alpha$, i.e., if 
\[
N-1-\alpha \geq \alpha
\]
or $\alpha \leq  (N-1)/2$.  In this case, the number of agents playing the given action will remain maximal regardless of the actions of the non-faulty agents.}
\end{example}

%and all players playing each action $a_i$ with $P(a=i)=1/3$ for $i=1,2,3$ is a mixed-strategy Nash equilibrium.\footnote{These are the only symmetric Nash equilibria. Depending on the value of $N$ there could also be other asymmetric Nash equilibria. For example, if $N=3$, then each player choosing a different action $a\in\{1,2,3\}$ is a Nash equilibrium, which is also $0$-robust.} We note that when all agents play the same action $a\in \{1,2,3\}$ this is a $\lfloor (N-1)/2\rfloor$-tolerant Nash equilibrium since if $\lfloor (N-1)/2\rfloor$ agents deviate to a different action, there will still be $\lceil (N-1)/2\rceil$  agents playing the given action and so this must still be maximal.
%However, the mixed strategy equilibrium is $0$-tolerant since if one agent is faulty and instead plays a pure strategy $a\in \{1,2,3\}$, then all other agents would profit from deviating to that same strategy.

Our definition of a $\alpha$-tolerant Nash equilibrium is slightly different from the way that fault tolerant equilibria are defined in \cite{Ka18, GR14}.  In those papers, fault tolerance is presented as a property of a given (normal) Nash equilibria.  Given a fixed Nash equilibria, it is defined to be tolerant to $\alpha$ faults if for 
for {\it every} subset of $\alpha$ players, the remaining $N-\alpha$ players would not unilaterally deviate from the given equilibrium actions regardless of the actions of the faulty players. In the following we will refer to such an equilibrium as being {\it tolerant to any set of $\alpha$ faults} to differentiate it for our definition 
of an $\alpha$-fault tolerant equilibrium.
Here, instead of defining fault tolerance as a property of a given focal equilibrium, we define it in terms of an equilibrium among the non-faulty players for a given and known number of faulty players.  Our motivation for adopting this different definition is primarily driven by our approach in the next section of 
looking at equilibrium existence through the lens of best response correspondences.  

Given that we are focusing on symmetric anonymous games, one might wonder if there is any difference  between the resulting equilibria under these two definitions.  Note that for the game in Example~\ref{ex:1} the equilibrium where all players choose the same action is also tolerant  to any $\alpha \leq (N-1)/2$ faults. More generally, for any game, given a Nash equilibrium that is tolerant to any set of $\alpha$ faults, we can map this into a $\alpha$-tolerant equilibrium by simply setting the strategies of the non-faulty players to be the same as the equilibrium strategies for any set of agents in the focal equilibrium. 

Next, we consider whether the converse of this statement holds.  In other words, if a game has an $\alpha$-tolerant equilibrium, does this always correspond to a Nash equilibrium that is tolerant to any set of $\alpha$ faults?   As one step in this direction, we have the following lemma.

\begin{lemma}\label{lem1}
For $\alpha \geq 1$, 
any $\alpha$-tolerant Nash equilibrium can be extended to a $0$-tolerant Nash equilibrium by specifying an 
appropriate set of strategies for the faulty players and keeping the strategies of the non-faulty players fixed.
\end{lemma}
\begin{IEEEproof}
To show that such a set of strategies for the faulty players exist, consider the sub-game between the faulty players, when the 
strategies of the non-faulty players are fixed to their values in the $\alpha$-tolerant Nash equilibrium, i.e., each faulty player now seeks to maximize their pay-off assuming that the non-faulty player's actions are fixed.
This is sub-game is also finite game and so a Nash equilibrium exists. Now consider a strategy profile in which the faulty players play their equilibrium strategies in the sub-game and the non-faulty players play their equilibrium strategies from  the $\alpha$-tolerant equilibrium.. This must be a Nash equilibrium of the original game.
\end{IEEEproof}

However, as the next example shows, given a $\alpha$-tolerant Nash equilibrium it may not be possible to extend it to a Nash equilibrium which is tolerant to 
any set of $\alpha$ faults in the sense of \cite{Ka18,GR14}.

\begin{example}\label{ex:2}
{\it Consider a game with $N=6$ players, where each player can choose one of three actions $A, B$ or $C$.  A player receives a pay-off of $1$ if it choses the same action as at least one other player and a pay-off of $0$, otherwise.  One $2$-tolerant Nash equilibrium if for two non-faulty players to choose action $A$ and the other two to choose action $B$.  Regardless of the action of the two faculty players, the non-faulty players would not want to unilaterally deviate.  

We can extend this to a $0$-tolerant Nash equilibrium by either having the two faulty players both play action $C$ or having them each choose one action from $\{A,B\}$, which need not be the same for the two agents.  In either case, it can be seen that the resulting Nash equilibrium is not tolerant to any set of two faults. For example, consider the Nash equilibrium where the two original faulty players chose action $C$ so that there are two agents choosing every action. Given this Nash equilibrium profile, suppose instead that one of the agents choosing action $A$ and one choosing action $B$ are faulty. If both of these agents change to action $C$, the non-faulty agents choosing actions $A$ and $B$ would also want to deviate, showing that this equilibrium is not tolerant to this set of $2$ faults.  Likewise, if we consider a Nash equilibrium where the two original faulty player both choose one action from $\{A,B\}$, then not there must be one of these actions which is chosen by 3 or fewer players.  If all but one of those players is faulty and switches to a different action, then the remaining player would also want to switch, again showing that this is not robust to $2$ faults.
}
\end{example}

Note also that there are other Nash equilibria for the game in  Example~\ref{ex:2}, that are tolerant to any set of $2$ faults in the sense of \cite{Ka18,GR14}. Namely, any equilibria where all 6 players choose the same action, which is in fact tolerant to any set of $4$ faults.

Lemma~\ref{lem1} also provides one way of checking for $\alpha$-tolerant Nash equilibria. Namely, one can look at all equilibria of a game and see if those are tolerant to faulty behavior by some set of $\alpha$ agents. Compared to \cite{Ka18,GR14}, we only need to find one set of $\alpha$ agents as opposed to checking this for every set.

The largest possible value of $\alpha$ in a game with $N$ players is $N-1$. A $N-1$-tolerant Nash equilibrium corresponds to a weakly dominant strategy for the non-faulty agent as it would want to continue playing the strategy regardless of the actions of the other $N-1$ players.  Moreover, since we are considering games with symmetric pay-offs, it must be that this strategy is weakly dominant for every player.  Hence, in this case, we can always extend the $N-1$-tolerant Nash equilibrium to a Nash equilibrium that is also a dominant strategy equilibrium.  Further, the resulting equilibrium will be tolerant to any set of $N-1$ faults. 

We next give a generalization of Lemma~\ref{lem1}, which shows that if a symmetric anonymous game  has a $\alpha$-tolerant Nash equilibrium, it also has an $\alpha -1$-tolerant equilibrium.\footnote{When a Nash equilibrium is tolerant to any set of $\alpha$ faults in the sense of \cite{Ka18,GR14}, it is straightforward to see that it is also tolerant to any set of $\alpha-1$ faults as this is just reducing the number of faults that can occur at the given equilibrium.  This idea underlies the defection index introduced in \cite{Ka18}, which is based on the largest such $\alpha$. In our case, this nesting is not immediate as when we vary $\alpha$ we are looking at equilibria of different games.}
\begin{lemma}\label{lem2}
For $\alpha \geq 1$, if a symmetric anonymous game has a $\alpha$-tolerant Nash equilibrium then it also has a $\alpha-1$-tolerant Nash equilibrium such that the strategies players $1,\ldots,N-\alpha$ are the same in both equilibria.
\end{lemma}
\begin{IEEEproof}
Assume a game has a $\alpha$-tolerant Nash equilibrium. To construct a $\alpha-1$-tolerant Nash equilibrium, let the non-faulty players $1,\ldots,N-\alpha$  use the same strategy as in the $\alpha$-tolerant Nash equilibrium. Since we are reducing the number of faulty players, these players will not have an incentive to deviate.  For  player $N-\alpha +1$, set its strategy to be the same as the strategy of player 1.  Since we are assuming anonymous, symmetric pay-offs, it follows that player $N-\alpha +1$ will also not have an incentive to deviate (since by assumption, player 1 does not have such an incentive when player $N-\alpha +1$ adopts this strategy).
\end{IEEEproof}

Note that this lemma provides an alternative proof of Lemma~\ref{lem1}. However, it is limited to symmetric anonymous games, while the proof of Lemma~\ref{lem1} applies more generally.

Following the proof of Lemma~\ref{lem2}, it follows that any $\alpha$-tolerant symmetric Nash equilibrium, can be extended to a symmetric $0$-tolerant Nash equilibrium. In this case it is also straightforward to see that this symmetric Nash equilibrium is tolerant to any set of $\alpha$ faults since there is no difference between every subset of $\alpha$ players in such an equilibrium.  In other words, for symmetric equilibria these two definitions of fault tolerance are equivalent.  
While as we saw in Example~\ref{ex:2}, with asymmetric equilibria, our notion of $\alpha$-fault tolerance is a weaker equilibrium concept.

\section{Best Response Correspondences and Equilibrium Existence}
\label{sec:Exist}

It is known there always exists a Nash equilibrium for a finite game~\cite{Na50}, i.e., there always exists a $0$-tolerant Nash equilibrium. A proof of this result comes from viewing a Nash equilibrium in terms of {\it best response correspondences}, where 
player $i$'s best response correspondence to the mixed strategy profile $\underline{\sigma}^{-i}$ of the other players is given by
\eq{\label{eq:BR}
BR^i(\underline{\sigma}) = \arg\max_{\sigma^i \in \mathcal{A}} U_i(\sigma^i,\underline{\sigma}^{-i}).
}
It then follows that  $\underline{\sigma}^*$ is a {\it Nash equilibrium} of $G$ if and only if for all $i$
\eq{\label{eq:BR-NE}
\sigma^{i,*} \in BR^{i}(\underline{\sigma}^{*}).
}

Let $BR(\underline{\sigma}) = (BR^i(\underline{\sigma}))_{i=1}^N$ be the {\it joint best response correspondence} which is a map from $\Delta(\mathcal{A})^N$ into $\Delta(\mathcal{A})^N$.
The common proof of existence of a Nash equilibrium then uses Kakutani's fixed-point theorem to show that this map has a fixed point 
\eq{\label{eq:fx1}
\underline{\sigma}^* \in BR^{i}(\underline{\sigma}^{*}),
}
which is equivalent to (\ref{eq:BR-NE}) holding for all $i$.

In this section, we seek to understand when this type of approach can be used to show the existence of the $\alpha$-tolerant Nash equilibrium.
Note that as shown by the next example, there are  symmetric anonymous games for which no $\alpha$-tolerant Nash equilibrium exists for any $\alpha >0$. This suggests that such an approach must put additional limits on the set of games under consideration.

\begin{example}[A congestion game] {\it Consider the following congestion game with $N=3$ players. Each player has a choice of two roads, $A$ or $B$ and receives a pay-off given by the negative total number of other players choosing the same road.
For example, if there are $2$ players choosing road $A$, then each receives a pay-off of $-2$.   It can be seen that the only $0$-tolerant Nash equilibrium of this game is for each player to use a mixed strategy in which they select each road with probability $1/2$.  Hence, from Lemma~\ref{lem1}, if this game has a $1$-tolerant Nash equilibrium, it must be one in which the 2 non-faulty players choose each road with probability $1/2$. But this can not be $1$-tolerant because if the one faulty player chooses road $A$, then a non-faulty player would receive an expected  pay-off of $1+ 1/2$ from choosing $A$ and an expected pay-off of $1/2$ from choosing $B$ and so it would want to deviate to choose road $B$, showing that a $1$-tolerant equilibrium does not exist.}
\end{example}

\subsection{Fault Tolerant Best Responses}

As with Nash equilibrium, we can again view $\alpha$-tolerant equilibria in terms of best response correspondences. Only now, these correspondences need to account for the behavior of the faulty agents.  First consider a fixed strategy profile of the non-faulty and faulty agents given by $(\underline{\sigma}^{(\alpha)},\underline{\tau}^{\alpha})$. A non-faulty agent $i$'s best response to this profile is then given by $BR^i(\underline{\sigma}^{(\alpha)},\underline{\tau}^{\alpha})$, where $BR^i$ is defined in  (\ref{eq:BR}).   Here, we are using the fact that once we fix the strategy profiles of the faulty and non-faulty agents to be $(\underline{\sigma}^{(\alpha)},\underline{\tau}^{\alpha})$, from the point-of-view of agent $i$'s best response, it makes no difference that some of the agents are faulty. It is simply best-responding to the given strategy profile.  We then define a non-faulty agent $i$'s 
{\it $\alpha$-tolerant best response correspondence} to a non-faulty action profile $\underline{\sigma}^{(\alpha)}$ as
\eq{\label{eq:aBR1}
BR^{i,\alpha}(\underline{\sigma}^{(\alpha)}) = \bigcap_{\underline{\tau}^{\alpha} \in T^{\alpha}} BR^i(\underline{\sigma}^{(\alpha)},\underline{\tau}^{\alpha}). 
}
In other words, if a non-faulty agent $i$ has a strategy that is in $BR^{i,\alpha}(\underline{\sigma}^\alpha)$, then it must be in $BR^i(\underline{\sigma}^\alpha,\underline{\tau}^{\alpha})$ for all possible choices of $\underline{\tau}^{\alpha} \in T^{\alpha}$, meaning that it is a best response for agent $i$ to chose this strategy regardless of the strategies of the faulty players.  

Using this definition, it follows that $\underline{\sigma}^{*(\alpha)}$ is a {\it $\alpha$-tolerant Nash equilibrium} if and only if for all non-faulty players $i$, 
\[
\sigma^{i,*(\alpha)} \in BR^{i,\alpha}(\underline{\sigma}^{*(\alpha)}).
\]

Let $BR^{\alpha}(\underline{\sigma}^{(\alpha)}) = (BR^i(\underline{\sigma}^{(\alpha)}))_{i=1}^{N-\alpha}$ be the {\it joint $\alpha$-tolerant best response correspondence} for the set of non-faulty players which maps $\Delta(\mathcal{A})^{N-\alpha}$ into $\Delta(\mathcal{A})^{N-\alpha}$.
The existence of an $\alpha$-tolerant Nash equilibrium then corresponds to this map having a fixed point
\eq{\label{eq:aBR}
\underline{\sigma}^{*(\alpha )} = BR^{\alpha}( \underline{\sigma}^{*(\alpha )}).
}

\subsection{Existence}

The following result shows that if the joint $\alpha$-tolerant best response is non-empty for all possible strategy profiles of the non-faulty agents, then such a fixed point exists.

\begin{theorem}\label{thm1}
If $BR^{\alpha}(\underline{\sigma}^{(\alpha)})$ is non-empty for all $\underline{\sigma}^{(\alpha)}$, then there exists an $\alpha-$tolerant Nash equilibrium. 
\end{theorem}
\begin{IEEEproof}
We prove this using Kakutani's fixed point theorem, which states that  a correspondence $F: X \mapsto 2^X$ has a fixed point if the following three properties hold:
\begin{enumerate}
\item $X$ is a non-empty, convex and compact susbset of $\mathbb R^N$,
\item $F(x)$ is non-empty, closed and convex for all $x\in X$,
\item $F$ is upper-hemi-continuous.
\end{enumerate}
The proof that every finite game has a $0$-tolerant Nash equilibrium follows from showing that these three requirements 
are all satisfied when $F$ is the joint best-response correspondence for finite game and 
$X = \Delta(\mathcal{A})^N$ is the set of all mixed strategy profiles across the agents. 

Here, we instead apply this for the joint $\alpha$-tolerant best response correspondence in (\ref{eq:aBR}). The domain of this correspondence is the set of all mixed strategy profiles of the non-faulty agents ($\Delta(\mathcal{A})^{N-\alpha}$), which is non-empty, convex and compact by the same reasoning as in the $0$-tolerant case.  Hence, to complete the proof, we only need to show the second and third properties.

For the second property, note that $BR^{\alpha}(\underline{\sigma}^{(\alpha)})$ is non-empty by assumption.  Next, note that for any fixed profile $(\underline{\sigma}^{(\alpha)},\underline{\tau}^{\alpha})$, $BR^i(\underline{\sigma}^{(\alpha)},\underline{\tau}^{\alpha})$ 
is simply the best response of player $i$ to this profile in the $0$-tolerant case. Since the underlying game is a finite game, this means that the above properties must hold for $BR^i(\underline{\sigma}^{(\alpha)},\underline{\tau}^{\alpha})$. In particular, this means that
$BR^i(\underline{\sigma}^{(\alpha)},\underline{\tau}^{\alpha})$ is compact and convex for every choice of $\underline{\tau}^{\alpha}$ and every choice of $\underline{\sigma}^{(\alpha)}$.  It then follows that $BR^{i,\alpha}(\underline{\sigma}^{(\alpha)})$ is convex as the intersection of closed convex sets is a closed convex set. Hence,  $BR^{\alpha}(\underline{\sigma}^{(\alpha)})$ is closed and convex as it is the Cartesian product of closed convex sets. Thus, the second property holds.

For the third property, first note that for a fixed $\underline{\tau}^{\alpha}$, we can view $BR^i(\underline{\sigma}^{(\alpha)},\underline{\tau}^{\alpha})$ as the best response correspondence of a non-faulty player $i$ to the profile $\underline{\sigma}^{(\alpha)}$ in a game 
$G'$ among the $N-\alpha$ non-faulty players, where the players' utilities in $G'$ are parameterized by $\underline{\tau}^{\alpha}$. As $G'$ is a finite game, it then follows that $BR^i(\underline{\sigma}^{(\alpha)},\underline{\tau}^{\alpha})$ is an upper-hemi-continuous function of $\underline{\sigma}^{(\alpha)}$  for every choice of $\underline{\tau}^{\alpha}$.  To complete the proof, we show that this implies that  $BR^{i,\alpha}(\underline{\sigma}^{(\alpha)})$ is also upper-hemi-continuous, which follows from Lemma~\ref{lem:uhc}, below.  Finally, if $BR^{i,\alpha}(\underline{\sigma}^{(\alpha)})$ is upper-hemi-continuous for each $i$, then it follows that $BR^{\alpha}(\underline{\sigma}^{(\alpha)})$  is also upper-hemi-continuous. This shows that the third required property holds and completes the proof.

\end{IEEEproof}

The proof of Theorem~\ref{thm1} uses the following lemma, whose proof is given in the Appendix~\ref{app:A}.

\begin{lemma}\label{lem:uhc}
Let $\Theta$, $X$ and $Y$ be compact subsets of $\mathbb R^n$.  For each $\theta \in \Theta$, let $f(\cdot,\theta): X \mapsto 2^Y$ be an upper-hemi-continuous correspondence with the property that $f(x,\theta)$ is non-empty and compact-valued for all $x\in X$.  Let $F:X\mapsto Y$ be another correspondence defined by 
\[
F(x) = \bigcap_{\theta \in \Theta} f(x,\theta).
\]
Then, if $F(x)$ is non-empty for all $x\in X$, it is also upper-hemi-continuous.
\end{lemma}

In the proof of Theorem~\ref{thm1}, we apply this lemma with $f$ being $BR^i(\underline{\sigma}^{(\alpha)},\underline{\tau}^{\alpha})$, $\theta$ being $\underline{\tau}^{\alpha}$ and $F$ being $BR^{i,\alpha}$.  

\subsection{Non-empty fault tolerant best responses}

The only restriction that Theorem~\ref{thm1} makes is that the joint $\alpha$-tolerant best response is non-empty for all strategy profiles of the non-faulty agents.  When $\alpha = 0$, this is always satisfied as the $0$-tolerant best response is simply the solution of the optimization a continuous function over a compact set, which must have a solution by the Weierstrass theorem.  Likewise, when $\alpha >0$, for any fixed $\underline{\tau}^{\alpha}$, $BR^i(\underline{\sigma}^{(\alpha)},\underline{\tau}^{\alpha})$ will be non-empty.  The issue is that the intersection of 
these correspondences over $\underline{\tau}^{\alpha}$ as in (\ref{eq:aBR1}) may be empty; i.e., there may not be a strategy for an agent $i$ that is a best response to every strategy of the faulty agents, as show in the next example.

\begin{example}
{\it We return to the popularity game in Example~\ref{ex:1}, which recall had a $\alpha$-tolerant Nash equilibrium for any $\alpha \leq (N-1)/2$ in which all non-faulty agents played the same action.  Suppose that $\alpha =1$ and that $N$ is an even number and larger than 3. Consider the following strategy profile of the non-faulty agents: half of them play action 1 and half play action 2. If the single faulty agent plays action 1, then the best response of every non-faulty player would be $\{1\}$.  Likewise, if the faulty player plays action 2, then the best response of every non-faulty player would be action $\{2\}$.  The intersection of these two sets is empty and so the $\alpha$-tolerant best response is empty for this strategy profile, meaning that Theorem~\ref{thm1} does not apply, even though this game has a $\alpha$-tolerant equilibrium. }
\end{example}

As an example where the condition in Theorem~\ref{thm1} is satisfied, consider a game in which the players have a weakly dominant strategy, then given any $\alpha \leq N-1$, this weakly dominant strategy will be in the $\alpha$-tolerant best response for every non-faulty player. Of course for such a game, Theorem~\ref{thm1} is not needed to see that an $\alpha$-tolerant equilibrium exists, the point here is just to show that the conditions in the this theorem may be satisfied.  One might hope that for game with "nearly" dominant strategies that this condition would also be satisfied, for example in a game with a equilibrium that is tolerant to $N-2$ faulty players, one fewer than in a dominant strategy.
The following example shows that this may not be the case.

\begin{example}\label{ex:match2}
{\it Consider a game similar to that in Example~\ref{ex:2}, in which players choose between 2 actions $\{A,B\}$ and receive a pay-off of $1$ if they choose the same action as any other player and a pay-off of $0$ otherwise.  A profile in which the non-faulty players choose the same action is $N-2$-tolerant.  Moreover, it can be seen that given $\alpha = N-2$ faulty players, a non-faulty player's $\alpha$-tolerant best response is non-empty for every pure strategy profile of the two non-faulty players, as matching the other non-faulty player's action is always a best response, regardless of the actions of the faulty players.  However, consider the mixed strategy profile in which the two non-faulty players are choosing each action with a non-zero probability.  In this case, if all of the dominant players choose the action $A$, then the best response of a non-faulty player would be to also choose action $A$, while a non-faulty player's best response would be to choose action $B$ if all the faulty players choose $B$. So, once again the $\alpha$-tolerant best response would be empty for this strategy profile.}
\end{example}

\subsection{Local fault tolerant best responses}

The previous examples suggest that the existence of a fault-tolerant equilibrium ioften depends on  "local" features of the fault tolerant best response correspondences as opposed to "global" features. With this in mind, suppose that we restrict 
$\underline{\sigma}^{\alpha}$ to be in any non-empty compact convex set $\mathcal{S} \subset \Delta(\mathcal{A})^{N-\alpha}$ and consider the correspondence $\hat{BR}^\alpha : \mathcal{S}\mapsto 2^\mathcal{S}$ defined by
\[
\hat{BR}^\alpha(\underline{\sigma}^{\alpha}) = BR^\alpha(\underline{\sigma}^{\alpha}) \cap \mathcal{S}.
\]
We then have the following corollary to Theorem~\ref{thm1}.
\begin{corollary}\label{cor1}
If $\hat{BR}^{\alpha}(\underline{\sigma}^{(\alpha)})$ is non-empty for all $\underline{\sigma}^{(\alpha)} \in \mathcal{S}$, then there exists an $\alpha-$tolerant Nash equilibrium in $\mathcal{S}$. 
\end{corollary}

In other words, for a $\alpha$-tolerant Nash equilibrium to exist, it is sufficient for the  $\alpha$-tolerant best response to be non-empty when restricted to 
some local neighborhood $\mathcal{S}$ of the equilibrium.  This is illustrated in the following example, which also shows that in some case the only set $\mathcal{S}$ that can be used are essentially ones that contain a single point corresponding to an $\alpha$-tolerant equilibrium. 
\begin{example}
{\it Consider a game in which the players choose one of two actions $\{A,B\}$ and receive a pay-off proportional to the total number of people choosing the same action as them (including themselves).  For any $\alpha \leq N/2$, this game has a $\alpha$-tolerant Nash equilibrium in which all players choose the same action. For a non-faulty player $i$, let $\sigma^i = (p^i_A,p^i_B)$ denote its mixed strategy. Define a set $\mathcal{S}_A$ of strategies of the non-faulty players in $\Delta({\mathcal A})^{N-\alpha}$ which satisfy
\eq{\label{eq:S}
\sum_{i=1}^{N-\alpha} p^i_A \geq \sum_{i=1}^{N-\alpha} p^i_B + \alpha.
}
Then note that for any $\underline{\sigma}^{\alpha} \in \mathcal{S}_A$, a non-faulty player's best response to any profile for the faulty agents will include the action $A$, since (\ref{eq:S}) ensures that the expected pay-off for $A$ will be at least as large as that of $B$.  Hence, $A$ is in the $\alpha$-tolerant best response for every $\underline{\sigma}^{\alpha} \in \mathcal{S}_A$ and so this choice of $\mathcal{S}_A$ satisfies Corollary~\ref{cor1}. 

In the same way it follows that if $\mathcal{S}_B$ was defined as  the set of strategies of the non-faulty players in $\Delta({\mathcal A})^{N-\alpha}$ which satisfy
\eq{\label{eq:S2}
\sum_{i=1}^{N-\alpha} p^i_A  + \alpha  \leq \sum_{i=1}^{N-\alpha} p^i_B,
}
then $B$ would be in the $\alpha$-tolerant best response for any profile in $\mathcal{S}_B$ and again Corollary~\ref{cor1} would be satisfied. For the set of profiles that do not satisfy (\ref{eq:S}) or (\ref{eq:S2}), it can be seen that the $\alpha$-tolerant best response will be empty.

Note also that as $\alpha$ increases, both $\mathcal{S}_A$ and $\mathcal{S}_B$ shrink and when $\alpha$ obtains the maximum value of $N/2$ (for $N$ even), then the only way that (\ref{eq:S}) holds is if $p^i_A = 1$ for all $i$ and the only way that (\ref{eq:S2}) holds if is $p^i_B = 1$ for all $i$.  In other words, at this choice of $\alpha$, $\mathcal{S}_A$ only contains the $\alpha$-tolerant equilibrium profile where all non-faulty players play $A$ and $\mathcal{S}_B$ only contains the profile where all non-faulty players play $B$ and for all other possible profiles of the non-faulty agents, the $\alpha$-tolerant best response is empty.
 }
\end{example}

When the set $\mathcal{S}$ only contains a single point, there is of course no -need to utilize a fixed point argument to see that an equilibrium exists, one can just verify that this point is an equilibrium directly by considering possible deviations. 

We also note that the formation index considered in \cite{Ka18} provides one way to construct a local set $\mathcal S$. This formation index  is defined as the smallest set of players in a given Nash equilibrium, such that when they all play an equilibrium action, it is a dominant strategy of every other player to play this action.  Suppose that an equilibrium has a formation index of $k$ and that there are at least $k$ non-faulty players. Then if 
we chose $\mathcal S$ to be all profiles in which the actions of first $k$ agents are fixed at their equilibrium value, it follows that the $\alpha$-tolerant best response of any remaining non-faulty player must include their action in the given equilibrium and so we can apply Corollary~\ref{cor1}.  However, as was  shown in \cite{Ka18}, such an equilibrium will be $\alpha$-tolerant for $\alpha \leq N-k$ and so if we chose the largest value of $\alpha$, this set $\mathcal S$ again contains just a single point, which is the equilibrium profile.

\subsection{A linear programing view}

In this section we view  $\alpha$-tolerant best responses correspondence through the view of the underlying linear programming problem.
For a given game with $K$ actions, let $\{a_1,\ldots,a_K\}$ denote the set of actions and let 
$\sigma^{i} = (p^i_k)_{k=1}^K$ denote a mixed strategy of agent $i$.
For a given choice of $\underline{\tau}^\alpha$ and $\underline{\sigma}^(\alpha)$, agent $i$'s best response is given by the following linear program
\eq{\label{eq:LP}
BR^i(\underline{\sigma}^{(\alpha)},\underline{\tau}^{\alpha}) = \arg\max_{\sigma^i \in \Delta({\mathcal A})} \sum_{k=1}^K p^i_k 
{\mathbb E} u(a_k,f(A^{-i})).
}
Here, the expectation is take with respect to the mixed strategies of the other non-faulty agents and well as the faulty agents.
Hence, varying these mixed strategies will change the cost coefficients (i.e., ${\mathbb E} u(a_k,f(A^{-i}))$)
in this linear program.  If the $\alpha$-tolerant best 
response of an agent is non-empty for a given profile $\underline{\sigma}^{(\alpha)}$, this means that there is a strategy that 
remains optimal for every set of cost-coefficients created by every choice of $\underline{\tau}^\alpha \in T^{\alpha}$. 
Conditions for when this occurs can be be derived by doing sensitivity analysis of the corresponding linear program.~\cite{BuLP}.

To get more insight, in the following we restrict our attention to games with $K=2$ actions.  Define a game with $\alpha$-faulty players as being {\it non-trivial} if changing the actions of the $\alpha$ faulty players change the gradient direction of the cost vector (e.g., this would not be the case in a game for which a player's pay-off did not depend on the actions of the other players).  
For a non-trivial game with faulty agents and 2 actions, it can be see that the only possible strategies in $BR^{i,\alpha}$ are extreme points of the constraint set
$\Delta({\mathcal A})$, which correspond  to a pure strategy for one of the two actions $a_1$ or $a_2$.  This is because mixed strategies would correspond to points in the interior of the face of the probability simplex connecting these two pure strategies. Such a point is an optimal solution to (\ref{eq:LP}) if and only if the gradient of the cost vector is orthogonal to the dominant face of the simplex and since the game is non-trivial, this can not be true for every choice of $\underline{\tau}^\alpha$. 
Next, suppose that the underlying game also does not have a weakly dominant strategy. We next argue that in this case, there must be a strategy profile 
$\underline{\sigma}^\alpha$ for which $BR^{i,\alpha}$ is empty.  Suppose that this was not the case. Then since the game does not have a dominant strategy, there must be some choice $\underline{\sigma}^\alpha_1$ for which $BR^{i,\alpha}(\underline{\sigma}^\alpha_1) = a_1$ and some choice $\underline{\sigma}^\alpha_2$ 
 for which $BR^{i,\alpha}(\underline{\sigma}^\alpha_2) = a_2$.   For $t\in [0,1]]$, let $\underline{\sigma}^\alpha(t)$ denote a continuous trajectory of mixed strategies starting with $\underline{\sigma}^\alpha(0) = \underline{\sigma}^\alpha_1$ and ending at $\underline{\sigma}^\alpha(1) = \underline{\sigma}^\alpha_2$.  Along this trajectory there must be at least one point where $BR^{i,\alpha}(\underline{\sigma}^\alpha(t))$ discontinuously changes from $a_1$ to $a_2$. However, this would violate the fact that $BR^{i,\alpha}(\underline{\sigma}^\alpha(t))$ is upper-hemi-continuous, contradicting our assumption that $BR^{i,\alpha}$ is always non-empty. 

In summary the above argument shows that for non-trivial games with 2 actions, that the only time the $\alpha$-tolerant best response is non-empty for all strategy profiles would be a game with a dominant strategy.

\section{Conclusions}
\label{sec:Conc}
We considered $\alpha$-tolerant Nash equilibria for finite symmetric anonymous games, which are equilibria that are robust to fault behavior by $\alpha$ players.  The notion of a $\alpha$-tolerant best response correspondence was introduced as a way to characterizing the existence of these equilibria view fixed point arguments, similar to Nash's approach for for showing equilibrium existence in finite games. Our results showed that when these correspondences are non-empty for every strategy profile of the non-faulty agents, then one can apply
Kakutani's fixed point theorem to show that $\alpha$-tolerant equilibrium exist.  For games with $N$ players, the $\alpha$ tolerant best response is non-empty for $\alpha =0$ (in which case it is the normal best response correspondence) and for $\alpha = N-1$ in games that are dominance solvable.  However, for
$0<\alpha <N-1$, we showed via examples that in many cases the correspondence will have empty values.  Using a linear programing argument for games with two actions, we showed that this will always be the case in games where the players do not have a dominant strategy.  We also showed that a fixed point exists for the $\alpha$-tolerant best responses by restricting them to a local neighborhood of a equilibrium profile, but in some cases the only possible neighborhood is one that only contains the equilibrium profile.  This shows that if general existence theorems for fault tolerant equilibria are to be found, it will require a different approach.

%We conclude with a few comments relating this back to some of the related work.  As we mentioned one line of 

\section{Acknowledgement}
Deepanshu Vasal would like to thank Arun Padakandla for helpful discussions.
\appendices
\section{Proof of Lemma~\ref{lem:uhc}}\label{app:A}
\begin{IEEEproof}
First note that since $f(x,\theta)$ is compact-valued for all $\theta$, it follows that $F(x)$ is also compact-valued.  To show that a compact-valued correspondence is upper-hemi-continuous, we use the following result:
\begin{lemma}\label{lem_h1}
For a non-empty and compact-valued 
correspondence $g:X\mapsto 2^Y$ is upper-hemi-continuous if and only if
for any $x^* \in X$ and any sequence $(x_t)$ in $X$ converging to $x^*$ and any sequence $(y_t)$ in $Y$ such that $y_t \in g(x_t)$ for each $t$, there is a $y^* \in g(x^*)$ and a subsequence of $(y_t)$ that converges to $y^*$.
\end{lemma}

Let  $x^*$, $(x_t)$ and $(y_t)$  be quantities as in lemma~\ref{lem_h1} for the correspondence $F$. To complete the proof we need to show that $y^* \in F(X^*)$ exists and that $(y_t)$ has a subsequence which converges to $y^*$.  Note that for any $t$, since $y_t \in F(x_t)$, it must be that $y_t \in f(x_t,\theta)$ for every $\theta \in \Theta$. For a given, $\theta \in \Theta$, since $f(x_t,\theta)$ is upper-hemi-continuous, then from Lemma~\ref{lem_h1}, it follows that there exists a sub-sequence of $(y_t)$ that converges to $y^*$, with $y^*\in f(x^*\theta)$. Re-define $(x_t)$ and $(y_t)$ to be just the elements corresponding to this convergent sub-sequence, so that $(x_t)$ converges to $x^*$ and $(y_t)$ converges to $y^*$.  Then, since $f(x_t,\theta)$ is upper-hemi-continuous for all $\theta \in \Theta$ it must be that $y^*\in f(x^*,\theta)$ for all $\theta$. Hence, $y^* \in F(x^*)$ as desired, completing the proof.
\end{IEEEproof}

\end{document}